\documentclass[12pt,preprint]{aastex}

\def\beq{\begin{equation}}
\def\enq{\end{equation}}  
\def\ba{\begin{eqnarray}}
\def\ea{\end{eqnarray}}
\def\<{\langle}
\def\>{\rangle}

\def\eps{\epsilon}

\begin{document}

\title{The stellar contribution to the extra-galactic background light
  and absorption of high-energy gamma-rays} \author{Soebur
  Razzaque\altaffilmark{1,2}, Charles D. Dermer\altaffilmark{1} and
  Justin D. Finke\altaffilmark{1,2}} \altaffiltext{1}{Space Science
  Division, Code 7653, U.S. Naval Research Laboratory, Washington, DC
  20375; srazzaque@ssd5.nrl.navy.mil} \altaffiltext{2}{National
  Research Council Research Associate}

\begin{abstract} TeV $\gamma$ rays from distant astrophysical sources
are attenuated due to electron-positron pair creation by interacting
with ultraviolet/optical to infrared photons which fill the universe
and are collectively known as the extra-galactic background light
(EBL).  We model the $\sim$0.1--10~eV starlight component of the EBL
derived from expressions for the stellar initial mass function, star
formation history of the universe, and wavelength-dependent absorption
of a large sample of galaxies in the local universe.  These models are
simultaneously fitted to the EBL data as well as to the data on the
stellar luminosity density in our local universe.  We find that the
models with modified Salpeter A initial mass function together with
Cole et al. (2001) or Hopkins \& Beacom (2006) star formation history
best represent available data.  Since no dust emission is included,
our calculated EBL models can be interpreted as the lower limits in
the $\sim$0.1--1~eV range.  We present simple analytic fits to the
best-fit EBL model evolving with redshift.  We then proceed to
calculate $\gamma$-ray opacities, and absorption of $\sim$10--300~GeV
$\gamma$-rays coming from different redshifts.  We discuss
implications of our results for the Fermi Gamma Ray Space Telescope
and ground-based Air Cherenkov Telescopes.  \end{abstract}

\keywords{stars: formation---stars: fundamental parameters---stars:
  luminosity function, mass function---dust, extinction---diffuse
  radiation---gamma rays: observations}

\section{Introduction}

Stars are the dominant sources of electromagnetic radiation in the
universe after the cosmic microwave background (see, e.g., Fukugita \&
Peebles 2004).  They emit radiation longward from ultraviolet to
infrared wavelengths.  However, photons with wavelength $\lesssim
2~\mu$m are highly absorbed by the dust in the host galaxies and only
a fraction of the radiation emitted by the stars escape to the
inter-galactic medium and form a diffuse background or EBL (see, e.g.,
Baldry \& Glazebrook 2003; Driver et al. 2008).  The dust in the host
galaxies, heated by the starlight, also radiate in the infrared
wavelengths and contribute to the EBL density at $\sim 10^{12}$ Hz.
It is the direct starlight component, $\lesssim 2~\mu$m or $\gtrsim
0.1$~eV, that affects the propagation of $\lesssim 5$~TeV
$\gamma$-rays from distant sources.  Indeed, the very soft spectral
energy distribution ($dN/dE \propto E^{-\Gamma}$) with $\Gamma \gtrsim
3$ observed from several TeV blazars at high redshift ($z\gtrsim 0.1$)
such as PKS 2155-304 (Aharonian et al. 2005), H 2356-309 (Aharonian et
al. 2006a); 1ES 1218+304 (Albert et al. 2006); 1ES 1101-232 (Aharonian
et al. 2006b); 0347-121 (Aharonian et al. 2007), 1ES 1011+496 (Albert
et al. 2007) and 3C 279 (Albert et al. 2008), and their cutoff at
$\gtrsim 1$~TeV are hints that high energy $\gamma$ rays from these
sources are absorbed by the EBL UV/optical photons (Persic \& de
Angelis 2008).  Lower energy ($<$TeV) $\gamma$-rays from high redshift
sources such as gamma-ray bursts (GRBs) and blazars can also probe the
EBL starlight component.

Calculation of the opacity of the universe to $\gamma$-rays by
$\gamma\gamma \to e^+e^-$ process dates back to Nishikov (1961),
followed by Gould \& Shr\'eder (1966) and Fazio \& Stecker (1970).
More recently Malkan \& Stecker (1998, 2001); Primack et al. (1999);
Kneiske, Mannheim \& Hartmann (2002); Kneiske et al. (2004); Primack,
Bullock and Somerville (2005); Stecker, Malkan \& Scully (2006)
calculated EBL models adopting either a phenomenological approach or
Monte Carlo galaxy formation code.  These models trace the general
trend of the data, which may be fitted with a combination of two or
more modified blackbody spectra for its two distinct peaks at the
infrared and optical wavebands (Dermer 2007).  Significant uncertainty
in data and large dispersion among models led to an indirect method to
constrain the EBL, namely by estimating change in spectral slope from
distant TeV blazars due to $\gamma\gamma$ absorption (Stecker \& de
Jager 1993; Stanev \& Franceschini 1998; Mazin \& Raue 2007). However,
such a method generally does not include possible absorption at the
source (see, e.g., Reimer 2007) and presumes a source spectrum.

In this paper, we build models of the EBL starlight component
($\sim$0.1--10~eV) directly from the stellar thermal surface
radiation.  Emission from an individual star during its main-sequence
lifetime is well approximated as a blackbody with a mass-dependent
temperature.  The post-main-sequence lifetime of a star is very short
compared to its main-sequence lifetime and their contribution at the
UV-optical wave bands is not significant.  They can, however,
contribute significantly to longer wavelengths due to their increased
luminosity in the post-main-sequence phase (Finke et al., in
preparation).  Only a small fraction of the stars with mass $\gtrsim
8M_\odot$ produce supernovae and even a smaller fraction produce GRBs.
Emission from these sources dominate the diffuse MeV background
(Watanabe et al. 1999; Ruiz-Lapuente, Cass\'e \& Vangioni-Flam 2001;
see, however, Strigari et al. 2005; Inoue, Totani \& Ueda 2008).
Emission from quasars and AGNs, on the other hand, dominate the
diffuse X-ray background (Mushotzky et al. 2000).  Contributions from
all these sources add only a small fraction to the total cosmic
electromagnetic energy density in the $\sim$ 0.1--10~eV range, and we
also ignore that.  The estimated lifetimes of individual stars depend
on their masses and the assumed cosmology, which is the standard
$\Lambda$CDM with ($h, ~\Omega_m, ~\Omega_\Lambda$) = (0.7,~0.3,~0.7),
and the Hubble constant $H_0 = 70 h_{0.7}$~km~s$^{-1}$~Mpc$^{-1}$.
Summing over contributions from stars of all masses formed in the
history of the universe then gives us the diffuse emission or EBL.

A sum over contributions from individual stars radiating at a given
redshift corresponds to the luminosity density or the stellar energy
emissivity of the universe at that redshift.  The initial mass
function (IMF), which is the distribution of stars by mass, and the
star formation rate (SFR), which is the mass that forms stars per unit
comoving volume per unit time, are two uncertain but related
parameters in our calculation.  We form classes of models by choosing
different combinations of these parameters and compare, in the
UV-optical band, the luminosity density data of the local universe
found from the surveys of nearby galaxies.  The same models are then
compared with EBL data.  Note that there are no adjustable free
parameters in our calculation once we choose a particular model.

Finally we use one of our best-fit models to calculate the
$e^\pm$ pair production opacities in the $\sim$10--300~GeV energy
range at different redshifts.  These results are applicable to
high-energy emission from distant sources such as GRBs and blazars
detected by the currently operating Fermi Gamma Ray Space Telescope
and Air Cherenkov Telescopes such as HESS, MAGIC and VERITAS.

In Sec. \ref{sec:formalism} we outline the formalism of our method and
introduce different models in Sec. \ref{sec:SFR_models} upon which we
base our EBL calculation.  We report our results in
Sec. \ref{sec:results}, compare with results from previous authors as
well as calculate EBL evolution with redshift.  We discuss
implications of our results for $\gamma$-ray astronomy in
Sec. \ref{sec:implications}, calculating $\gamma\gamma \to e^+e^-$
absorption opacity ($\tau_{\gamma\gamma}$) for high energy $\gamma$
rays from sources at different redshifts.  Conclusions of the work for
EBL and $\gamma$ ray absorption are given in
Sec. \ref{sec:conclusions}.

\section{Formalism}
\label{sec:formalism}

The differential number density (per unit volume) of thermal blackbody
photons in the energy interval $\eps$ to $\eps+d\eps$ at a given
temperature $T$ is
\beq 
\frac{dN}{d\eps dV} = \frac{1}{\pi^2 (\hbar c)^3} 
\frac{\eps^2}{\exp(\eps/kT)-1}
\label{bb-photon-density}
\enq
and the total number of photons emitted per unit energy and time
intervals from a star of radius $R$ is given by
\ba
\frac{dN(\eps,M)}{d\eps dt} &=& \pi R^2 c ~\frac{dN}{d\eps dV} ~.
\label{bb-photon-perstar}
\ea
The radius and temperature of a star depend on its mass $M$, and can
be calculated in terms of the solar radius $R_\odot$, mass $M_\odot$
and temperature $T_\odot$ using various relations.  We use a fit to
the stellar mass-radius relation (Schmidt-Kaler 1982; Binney \&
Merrifield 1998) as
\ba
R/R_\odot = \cases{ (M/M_\odot)^{0.8} ~;~ 
0.1M_\odot \le M\le 10 M_\odot \cr
10^{9/20} (M/M_\odot)^{0.35} ~;~ 10 M_\odot < M \le 120 M_\odot ~.
}
\label{mass-radius}
\ea

The stellar mass-luminosity ratio is not well-known.  To the first
approximation, it is a single power law (SPL), given by
\begin{equation}
L/L_\odot = (M/M_\odot)^{3.6}\;.
\label{LoverLodot}
\end{equation}
A more detailed relation is calculated by Bressan et al. (1993) and a
simple broken power law (BPL) fit to that relation (Binney \&
Merrifield 1998) is given by
\ba
L/L_\odot =  f_{L} \cases{ (M/M_\odot)^{4.8} 
~;~ M < 2 M_\odot \cr
2^{13/10} (M/M_\odot)^{3.5} 
~;~ 2 M_\odot \le M \le 20 M_\odot \cr
2^{201/50} 5^{34/25} (M/M_\odot)^{2.14} 
~;~ M > 20 M_\odot ~.
}
\label{mass-luminosity}
\ea
Although there are uncertainties, depending on the metallicity e.g.,
in the mass-to-light ratio, we assume $f_{L}=1$ for our modeling
purposes.  Stars below 1$M_\odot$ and above 20$M_\odot$, in the
Bressan et al. (1993) model, produce much less light than the model
with a single power-law [equation~(\ref{LoverLodot})] for the same
masses.  We derive a stellar mass-temperature relation from the
luminosity $L=4\pi R^2 \sigma T^4$, where $\sigma$ is the
Stefan-Boltzmann constant.  For the SPL relation in
equation~(\ref{LoverLodot}) between $L-M$, the temperature is
\ba
T/T_\odot = \cases{
(M/M_\odot)^{1/2} ~;~ M\le 10M_\odot \cr 
10^{-9/40} (M/M_\odot)^{0.725} ~;~ M> 10M_\odot ~,
}
\label{MT-relation}
\ea
following the break in equation~(\ref{mass-radius}).  For the BPL
relation of $L-M$ in equation~(\ref{mass-luminosity}), the $T-M$
relation is
\ba
T/T_\odot &=& f_{L}^{1/4} 
\cases{
(M/M_\odot)^{0.8} ~;~ 0.1M_\odot \le M < 2 M_\odot \cr
2^{13/40} (M/M_\odot)^{0.475} ~;~ 
2 M_\odot \le M \le 10 M_\odot \cr
2^{1/10} 5^{-9/40} (M/M_\odot)^{0.7} ~;~ 
10 M_\odot < M \le 20 M_\odot \cr
2^{39/50} 5^{23/200} (M/M_\odot)^{0.36} ~;~ M > 20 M_\odot ~.
}
\label{mass-temperature}
\ea
For reference, $M_\odot = 1.99\times 10^{33}$~g, $L_\odot =
3.846\times 10^{33}$~ergs~s$^{-1}$ and $T_\odot = 5777$~K are solar
mass, luminosity, and temperature respectively.

\begin{figure}
\plotone{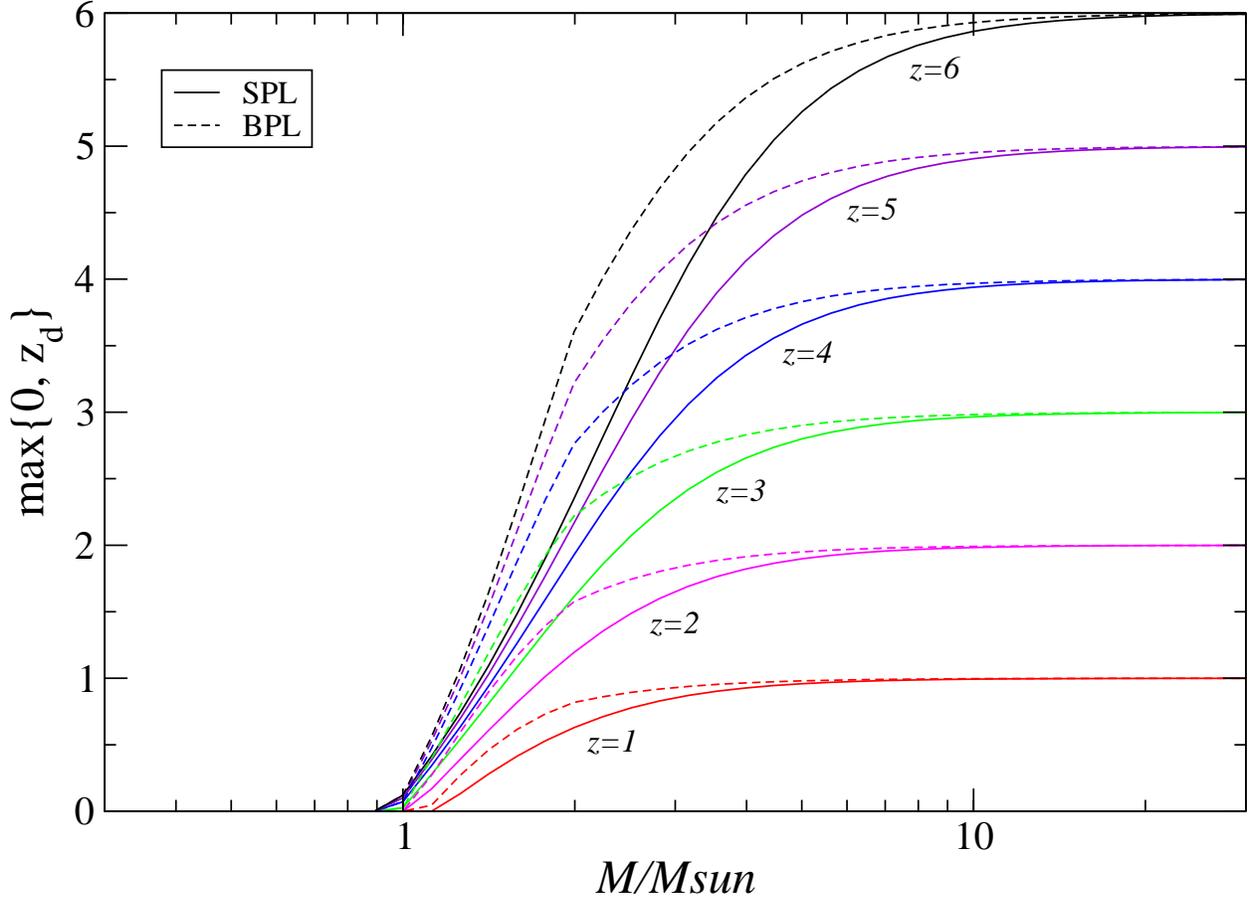}
\caption{Plot of the redshift (${\rm max}\{0,~z_{\rm d} (M,z)\}$) at
  which a star of mass $M$ which was born at redshift $z=$1 -- 6
  evolves off the main sequence using equation~(\ref{birth-death-z})
  and assuming the standard (0.7, 0.3, 0.7) $\Lambda$CDM cosmology.
  The two sets of curves correspond to SPL (dashed) and BPL (dotted)
  $L-M$ relation, respectively, in equations~(\ref{LoverLodot}) and
  (\ref{mass-luminosity}).}
\label{fig:zdeath}
\end{figure}

One needs to take cosmology into account to calculate the total number
of photons emitted from a star over cosmic time which was born at a
past epoch.  The relationship between the cosmic time and redshift is
given by
\beq
\left( dt/dz \right)^{-1} = 
-H_0(1+z) \sqrt{\Omega_m (1+z)^3 + \Omega_{\Lambda}} ~.
\label{cosmology}
\enq
The main sequence lifetime of a star with mass $M$ is 
\begin{equation}
t_\star \approx
t_\odot (M/M_\odot)/(L/L_\odot)\;,
\label{todot}
\end{equation}
where $t_\odot \approx 11$~Gyr is the lifetime of the Sun.  If a star
of mass $M$ was born at a redshift $z$ then the redshift $z_{\rm d}
(M)$ at which it had evolved off the main sequence can be found from
the inverse of the relation $t_\star (M) = \int_{z_{\rm d} (M)}^z dz'
\left| dt/dz' \right|$ as
\ba
z_{\rm d} (M,z) = -1 + 
\left( -(\Omega_\Lambda/\Omega_m)~
{\rm sech} \left[ (3/2) H_0 t_\star \sqrt{\Omega_\Lambda} 
\right. \right. \nonumber \\ + \left. \left.
\tanh^{-1} \sqrt{1+ (\Omega_m/\Omega_\Lambda) (1+z)^3 } 
\right]^2 \right)^{1/3} \;,
\label{birth-death-z}
\ea
We have plotted this redshift in Fig.~\ref{fig:zdeath} for different
$M$ and $z$ and setting it equal to zero if $z_{\rm d} (M) <0$.  After
calculating $z_{\rm d}$ for a given $(z,M)$ combination, we
back-calculate $z$ as a check.  The results are identical.  Note that
high mass stars $\gtrsim 10M_\odot$ evolve off the main sequence
almost at the same redshift ($z\approx 1$ -- 5) they were formed.  On
the contrary, a star with mass $\sim 1M_\odot$ lives almost a Hubble
time.  Note that $z_{\rm d}$, following $t_\star$, depends on the
$L-M$ relation.  For the curves in Fig.~\ref{fig:zdeath} we used both
the SPL (dashed lines) and BPL (dotted lines) $L-M$ relations given,
respectively, by equations~(\ref{LoverLodot}) and
(\ref{mass-luminosity}).

Equation~(\ref{todot}) represents an underlying source of uncertainty
for high-mass stars and/or high redshift stars.  Our naive estimate of
stellar lifetime and luminosity are based on the model of the Sun, and
assuming that stars are perfect blackbodies. Estimates of the Sun's
age are based on the total amount of fusion material ($4{\rm H}^+ \to
{\rm He}^{++}$) in the core.  This amounts to about $10\%$ of the
Sun's mass.  The estimates for high mass stars or stars with different
metallicities are less precise.  High mass stars often have winds
which may reduce their net radiative output, however, they can also be
approximated as perfect blackbodies because of their higly ionized
surface.  Since the fraction of high-mass stars is small, we neglect
corrections to equation~(\ref{todot}) in this study.

The integrated number of photons emitted by a star from its birth at
redshift $z$ to the present epoch, in the energy interval $d\eps =
d\eps'/(1+z')$, can be calculated using
equations~(\ref{bb-photon-perstar}), (\ref{mass-radius}),
(\ref{LoverLodot}) or (\ref{mass-luminosity}), (\ref{MT-relation}) or
(\ref{mass-temperature}) and (\ref{birth-death-z}) as
\ba
\frac{dN(\eps,M)}{d\eps} = 
\int_{{\rm max}\{0,~z_{\rm d} (M,z')\}}^z dz' 
\left| \frac{dt}{dz'} \right| 
\frac{dN(\eps',M)}{d\eps' dt} (1+z') ~,~
\label{total-perstar}
\ea
where the lower limit of the integration is set to zero if the star
has survived to the present epoch.

The number of stars formed at a redshift $z$ depends on the initial
mass function and star formation rate, both of which are important
sources of uncertainty in our calculation and are discussed shortly.
We assume a universal IMF which is normalized between mass ($M_{\rm
  min}$ -- $M_{\rm max}$) = (0.1 -- 120)$M_\odot$ as ${\cal N}^{-1} =
\int_{M_{\rm min}}^{M_{\rm max}} dM (dN/dM)M$.  The final number of
photons reaching us from all stars created in the past, per unit
energy interval, amounts to weighting equation~(\ref{total-perstar})
by the normalized IMF, ${\cal N}(dN/dM)$, and integrating over stellar
mass (to take into account photons created by stars of all mass) and,
after weighting by the SFR $\psi (z)$ in units of $M_\odot$~yr$^{-1}$
~Mpc$^{-3}$, integrating over redshift $z$ (to take into account stars
created at all epochs).  The spectral stellar radiation density is
therefore given by
\ba
\frac{dN(\eps, z=0)}{d\eps dV} = 
{\cal N} \int_{z=0}^\infty dz'' 
\left| \frac{dt}{dz''} \right| \psi(z'') 
\int_{M_{\rm min}}^{M_{\rm max}} dM \left( \frac{dN}{dM} \right) 
\nonumber \\ \times 
\int_{{\rm max} \{0,~z_{\rm d} (M,z')\}}^{z''} dz' 
\left| \frac{dt}{dz'} \right| f_{\rm esc} (\eps') 
\frac{dN(\eps',M)}{d\eps' dt} (1+z').~~~~
\label{final-photon}
\ea
Here $dN(\eps',M)/d\eps' dt$ is given by
equation~(\ref{bb-photon-perstar}) with the substituton $\eps' = \eps
(1+z')$, and $f_{\rm esc} (\eps')$ is the escape fraction of photons
from the host galaxy, considered further in the next section.  Note
that there is no free parameters in our model once we chose particular
models of SFR, IMF and $L-M$ relation.

Equation (\ref{final-photon}) can be converted to the EBL energy
density $\eps u_\eps$ (e.g., in units of ergs~cm$^{-3}$) by
multiplying with $\eps^2$ or to intensity $\eps I_\eps$ (e.g., in
units of W~m$^{-2}$~sr$^{-1}$) by multiplying with $\eps^2 c/4\pi$.
The differential photon number density in
equation~(\ref{final-photon}) or EBL energy density measured at
present from a past epoch ($z=z_1$) can be transformed to the past
epoch, i.e. comoving EBL density, with $\eps_1 = \eps (1+z_1)$ and the
comoving volume $V_1 = V/(1+z_1)^3$ as
\ba
\frac{dN(\eps_1, z_1)}{d\eps_1 dV_1} &=& 
(1+z_1)^2 \frac{dN(\eps, z=z_1)}{d\eps dV}
\nonumber \\
\eps_1 u_{\eps_1} &=& (1+z_1)^4 \eps^2 \frac{dN(\eps, z=z_1)}{d\eps dV}.
\label{transformations}
\ea
Note that we use $z=z_1$ as the lower limit of the outer integration
over redshift in equation~(\ref{final-photon}) to calculate $dN(\eps,
z=z_1) /d\eps dV$.

\subsection{Energy output by the local universe}
\label{subsec:luminosity_density}

The cosmic energy output or luminosity density $\eps L_\eps$ (e.g. in
units of W~Mpc$^{-3}$) in starlight is generally found from the galaxy
counts in the local universe (see, e.g., Baldry \& Glazebrook 2003;
Driver et al. 2008).  This is equivalent to summing over energy output
by individual stars which were born at a past epoch but still
radiating today ($z=z_1\approx 0$), and we calculate the comoving
luminosity density by modifying equation~(\ref{final-photon}) as
\ba 
\eps_1 L_{\eps_1} &=& \eps^2 (1+z_1)^5 {\cal N} f_{\rm esc} (\eps_1) 
\int_{M_{\rm min}}^{M_{\rm max}} dM \left( \frac{dN}{dM} \right)
\frac{dN(\eps_1,M)}{d\eps_1 dt}
\nonumber \\  && \times 
\int_{z_1\approx 0}^{z_{\rm b} (M,z_1)} dz'' 
\left| \frac{dt}{dz''} \right|
\psi(z'') ~.
\label{local-power}
\ea
Here $z_{\rm b} (M,z_1)$ is the maximum redshift a star may born,
within comoving Hubble time $1/H_0(1+z_1)$, to contribute to the local
power depending on its mass.  This can be calculated from
equation~(\ref{birth-death-z}) by switching the sign of the second
term inside the square brackets.

Note that an integration of the luminosity density, the comoving
quantity in equation~(\ref{local-power}) divided by $(1+z_1)^4$ and
multiplying with it after integration, over cosmic time also leads to
the comoving EBL energy density, an approach advocated by Salamon \&
Stecker (1998) and by Dwek et al. (1998).  This amounts to adding up
luminosity densities from all past epochs measured at a particular
epoch to calculate the EBL.  Both this approach and our formalism
given by equations~(\ref{final-photon}) \& (\ref{transformations})
produce the same EBL result as we have tested at different redshifts.

\section{Models of Star formation and dust attenuation}
\label{sec:SFR_models}

The star formation history of the universe and the number of stars
formed at a given epoch are not independent.  A combination of both
are needed to fit the SFR data (see, e.g., Madau et al. 1996; Cole et
al. 2001; Hopkins \& Beacom 2006) and the luminosity density data from
local galaxy surveys (see, e.g., Baldry \& Glazebrook 2003; Driver et
al. 2008).  The classic Salpeter (1955) IMF is still preferred by
astronomers with some modifications.  The modified ``Salpeter A''
model IMF is $dN/dM \propto M^{-\kappa}$ with $\kappa = 1.5$ below
$0.5M_\odot$ and 2.35 above $0.5M_\odot$.  Models by Scalo (1986;
1998) are in violation with a strong upper limit of $\kappa < 2.7$
above $1M_\odot$ as found by Baldry \& Glazebrook (2003).  We also use
the ``Baldry-Glazebrook'' IMF model with $\kappa = 1.5$ below
$0.5M_\odot$ and $\kappa = 2.2$ above $0.5M_\odot$.

Cole et al. (2001) found a parametric form of the SFR given by
$\psi(z) = h(a+bz)/[1+(z/c)^d]$, with parameters ($a, b, c, d$) =
(0.0166, 0.1848, 1.9474, 2.6316).  Hopkins \& Beacom (2006) also used
this parametric form, as well as their own piecewise fit in the form
$\sim 10^{a'} (1+z)^{b'}$, along with the ``Salpeter A'' and
``Baldry-Glazebrook'' IMF to fit SFR data.  The best-fit parameters
they found are given in their Tables 1 and 2.  We define five models
of different IMF and SFR combinations as:

\begin{itemize}
\item {\em Model A:} Cole et al. (2001) SFR and Salpeter A IMF
\item {\em Model B:} Cole et al. (2001) SFR formula fitted by Hopkins \&
  Beacom (2006) with Salpeter A IMF
\item {\em Model C:} Cole et al. (2001) SFR formula fitted by Hopkins \&
  Beacom (2006) with Baldry-Glazebrook IMF
\item {\em Model D:} Hopkins \& Beacom (2006) SFR with Salpeter A IMF
\item {\em Model E:} Hopkins \& Beacom (2006) SFR with
  Baldry-Glazebrook IMF
\end{itemize}

\begin{figure}
\plotone{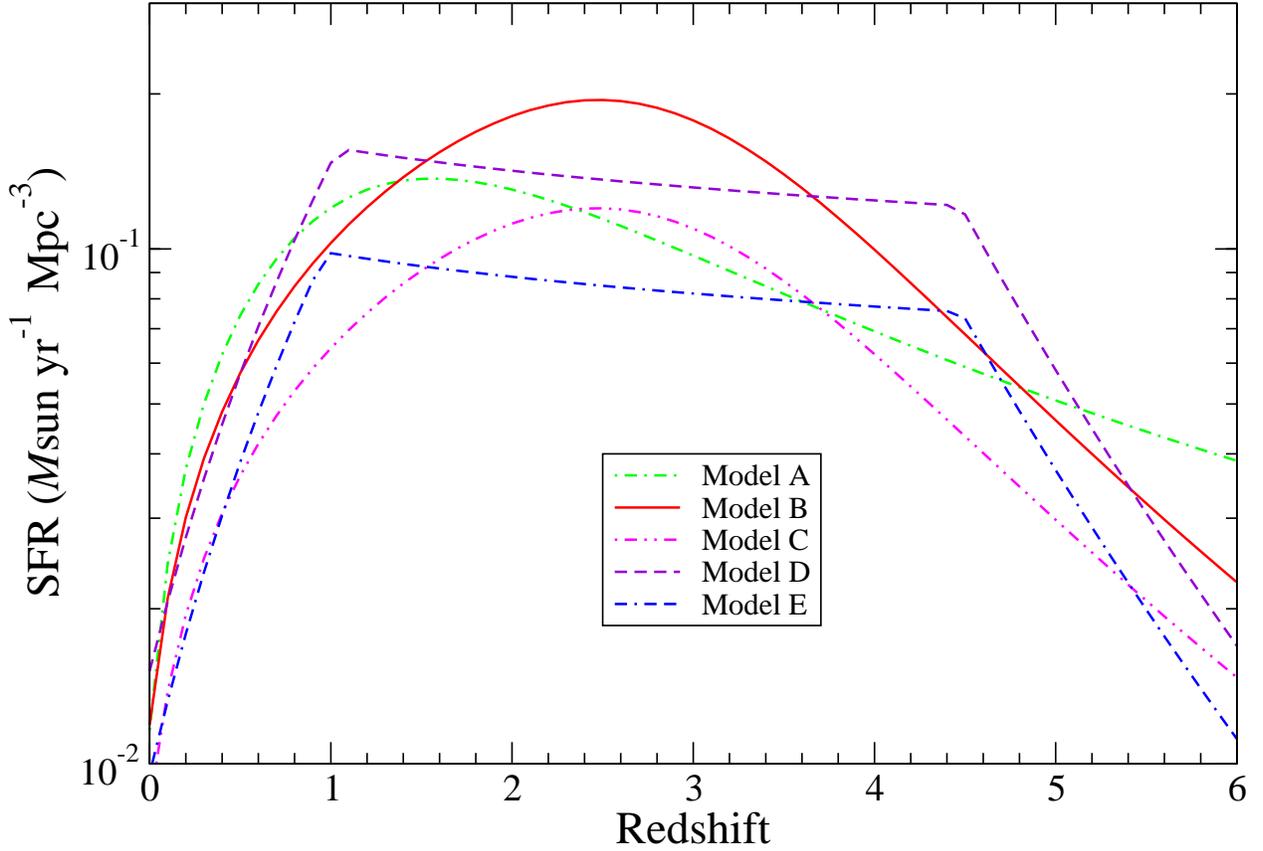}
\caption{Models of SFR combined with specific IMF.  {\em Model A}
  correspond to Cole et al. (2001) SFR with {\em Salpeter A} IMF, {\em
    Model B} correspond to Cole et al. (2001) SFR formula fitted by
  Hopkins \& Beacom (2006) with {\em Salpeter A} IMF, {\em Model C}
  correspond to Cole et al. (2001) SFR formula fitted by Hopkins \&
  Beacom (2006) with {\em Baldry-Glazebrook} IMF, {\em Model D}
  correspond to Hopkins \& Beacom (2006) SFR with {\em Salpeter A} IMF
  and {\em Model E} correspond to Hopkins \& Beacom (2006) SFR with
  {\em Salpeter A} IMF.}
\label{fig:sfr}
\end{figure}

We plot these models in Fig.~\ref{fig:sfr}.  Note that there is a
significant amount of uncertainty among the models even at $z\sim 0$.
Also, at the highest redshifts the SFR models may be underestimating
the true rate (see, e.g., Faucher-Gigu{\`e}re et al. 2008).  Most of
the uncertainty comes from the dust in the host galaxies that absorbs
stellar emission and reemits it into infrared wavebands, forming the
low energy, $<0.1$~eV, part of the EBL.  A precise model of the
fraction of starlight which directly escapes the host galaxy is still
missing.  Such a model would depend on the galaxy types and their
orientations as well as on redshift.

Bearing in mind the uncertainties in dust absorption models discussed
above, we adopt a model recently developed by Driver et al.\ (2008),
who have calculated the averaged photon escape fraction $f_{\rm esc}
(\lambda)$ from observations of 10,000 nearby galaxies convolved with
galactic dust models.  We fit their results with four segments as
\ba f_{\rm esc} (\lambda) = \cases{
0.688+0.556~{\rm log}\lambda ~;~ 
\lambda \le 0.165 \cr
0.151-0.136~{\rm log}\lambda ~;~
0.165 < \lambda \le 0.22 \cr
1.0+1.148~{\rm log}\lambda ~;~
0.22 < \lambda \le 0.422 \cr
0.728+0.422~{\rm log}\lambda ~;~
\lambda > 0.422\;,
}
\label{photon-escape}
\ea
with $\lambda$ in microns, and assume it to be universal or
independent of redshift.  The escape fraction $f_{\rm esc} \approx 0$
above $\approx 10$~eV as photons above this energy are absorbed by
galactic gas.  A feature of this model (Driver et al. 2008) is that
the total amount of stellar energy absorbed by dust, in our local
universe, is equivalent to the observed total luminosity density in
infrared photons.  The emission from luminous infrared galaxies, some
of which may have an AGN core, is an added correction to the infrared
EBL (see also Kneiske et al. 2004).

\section{Results}
\label{sec:results}

\begin{figure}
\plottwo{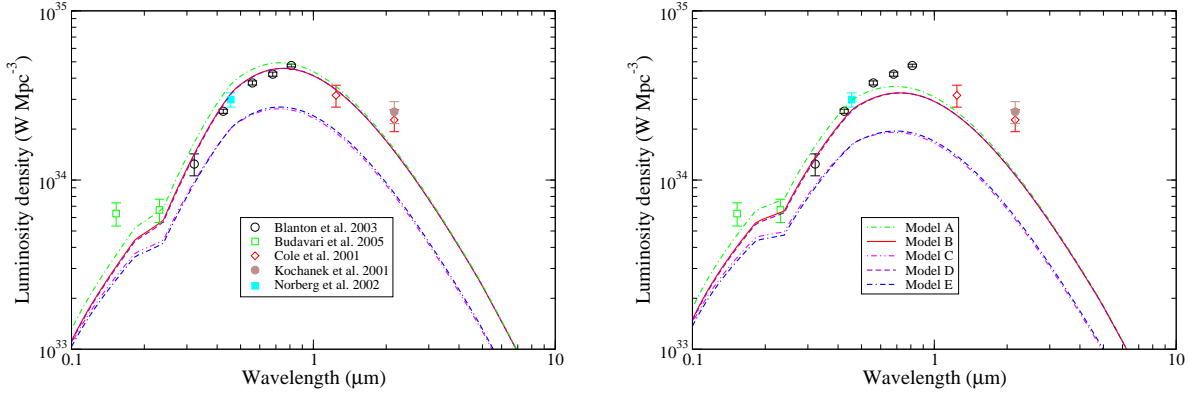}{f4b.eps}
\caption{Total energy output or luminosity density, $\eps_1
  L_{\eps_1}$ from equation (\ref{local-power}), in direct starlight
  by our local ($z_1= 0.1$) universe.  The data points are from {\em
    SDSS}, {\em 2MASS}, {\em 2dFGRS} and {\em GALEX} surveys of local
  galaxies.  The smooth lines are our calculations using
  equation~(\ref{local-power}) and based on SFR+IMF models A--E described
  in Section \ref{sec:SFR_models}.  The left and right panels
  correspond to the SPL and BPL models of $L-M$ relation,
  respectively, given by equations~(\ref{LoverLodot}) and
  (\ref{mass-luminosity}).}
\label{fig:lumdensity}
\end{figure}

\begin{figure}
\plottwo{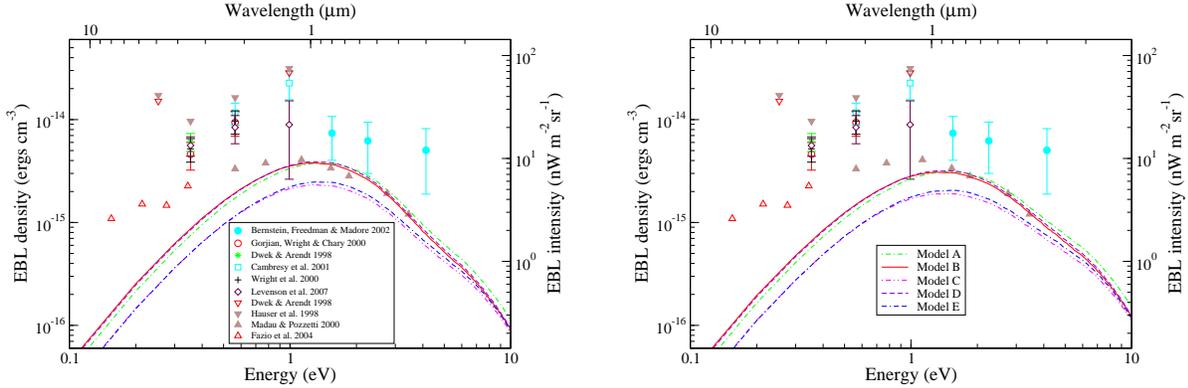}{f5b.eps} 
\caption{EBL energy density, $\eps u_\eps = \eps^2 dN(\eps,z=0)/d\eps
dV$ from equation (\ref{final-photon}), from direct starlight in our
local universe.  The smooth lines are calculated using
equation~(\ref{final-photon}) and correspond to SFR+IMF models A--E as
defined in Section \ref{sec:SFR_models}.  The data points with
errorbars, and lower (triangles) and upper (inverted triangles) limits
are described in Sec.~\ref{sec:results}.  {\em Left panel}--- models
with $L-M$ relation (SPL) in equation (\ref{LoverLodot}).  {\em Right
panel}--- models with $L-M$ relation (BPL) in equation
(\ref{mass-luminosity}).  }
\label{fig:EBL}
\end{figure}

We use Models A -- E for the SFR$+$IMF combinations and
equations~(\ref{final-photon}) and (\ref{local-power}), respectively,
to calculate the EBL at $z=0$ and luminosity density at $z=0.1$ in
starlight component, and compare with data.  The results, with
numerical integrations carried out by the multidimensional adaptive
Monte Carlo code Vegas (Lepage 1978; 1980), are plotted in
Figs.~\ref{fig:lumdensity} and \ref{fig:EBL} for the SPL (left panels)
and BPL (right panels) models of the $L-M$ relation.

To plot the luminosity density data points, in
Fig.~\ref{fig:lumdensity}, we took the AB magnitude data $m_{AB}
\equiv j+2.5~{\rm log}~h$ with $j=-2.5~{\rm log}(L_\eps /{\rm W
~Hz^{-1} Mpc^{-3}}) + 34.1$ and converted them to luminosity density
as $\eps L_\eps = h(c/\lambda) 10^{13.64-m_{AB}/2.5}$~W Mpc$^{-3}$.
The factor 34.1 comes from $2.5~{\rm log}(4.345\times 10^{13})$, where
$4.345\times 10^{13}$ ~W~Hz$^{-1}$ Mpc$^{-3}$ is the minimum of the
absolute AB magnitude scale.  Note that all data points are not
measured at the same redshift or corrected to $z=0$ as they come from
different surveys: The Sloan Digital Sky Survey ({\em SDSS}), the Two
Micron All Sky Survey ({\em 2MASS}), the Two-Degree Field Galaxy
Redshift Survey ({\em 2dFGRS}) and the Galaxy Evolution Explorer ({\em
GALEX}).  The data points by Budavari et al. (2005) are calculated for
redshift 0.07 -- 0.13 and the mean redshift is 0.1 for the data points
by Blanton et al. (2003).  The data point by Norberg et al. (2002) and
those by Cole et al. (2001) are corrected to $z=0$.  The redshift for
Kochanek et al. (2001) data point is 0.03.  Note that we plotted our
luminosity density models A--E for redshift 0.1 to be consistent with
most data points.

The three EBL data points in Fig.~\ref{fig:EBL}, in the UV band, are
calculated by Bernstein, Freedman \& Madore (2002) using measurements
from the Hubble Space Telescope ({\em HST}).  However, revised
(Mattila 2003; Bernstein, Freedman \& Madore 2005) and re-revised
(Bernstein 2007) estimates have put substantial uncertainty on these
data points.  The rest of the data points come from analyzing
measurements by the Diffuse Infrared Background Experiment ({\em
DIRBE}) onboard the Cosmic Background Explorer ({\em COBE}) satellite
by different authors (Dwek \& Arendt 1998a; Gorjian, Wright \& Chary
2000; Wright \& Reese 2000; Cambr\'esy et al. 2001 and Levenson,
Wright \& Johnson 2007).  This is the reason for multiple data points
at the same energies.  Note that the $\sim 1$~eV (1.25~$\mu$m) data
point by Levenson, Wright \& Johnson (2007) is really a $1\sigma$
limit.  We plot only those upper limits (Dwek \& Arendt 1998a; Hauser
et al. 1998) and lower limits (Madau \& Pozzetti 2000; Fazio et
al. 2004) which are directly derived from the {\em DIRBE} and {\em
HST} data.

Given a combination of mixed data points, Models A, B and D fare well
reproducing the luminosity density data in Fig.~\ref{fig:lumdensity}
with the SPL model of $L-M$ relation (left panel) given by
equation~(\ref{LoverLodot}).  These models miss the two data points at
the smallest and largest (within $2\sigma$) wavelengths, however.  In
case of the BPL model of $L-M$ relation (right panel) given by
equation~(\ref{mass-luminosity}), Models A, B and D agree with two
small wavelength data points better than the SPL model, but miss most
data points at longer wavelengths.  Models C and E are systematically
lower than all data points for both the SPL and BPL cases.  The
integrated total energy output by stars, using Model B (SPL), is
$1.2\times 10^{35}$~W~Mpc$^{-3}$ before dust absorption and $0.7\times
10^{35}$~W~Mpc$^{-3}$ after dust absorption.  As for comparison,
Driver et al. (2008) calculated these values as $1.6\times
10^{35}$~W~Mpc$^{-3}$ and $0.9\times 10^{35}$~W~Mpc$^{-3}$,
respectively.

Comparing with EBL data in Fig.~\ref{fig:EBL}, we find again that
Models A, B and D with the SPL $L-M$ relation (left panel) represent
the data points better than all other models in both SPL and BPL
cases.  Models A, B and D in the BPL case are consitently lower, below
2~eV, than the SPL case, a reflection of the trend observed in the
luminosity density plots as well.  Models C and E in both the SPL and
BPL cases (right panel) are below the {\em HST} lower limits (Madau \&
Pozzetti 2000).  None of our models are able to reproduce EBL data
below 1~eV hinting that an extra component is required for modeling
infrared data.

Models A, B and D involve Salpeter A IMF which predicts more stars
forming below $\sim 2M_\odot$ than the Baldry \& Glazebrook (2003) IMF
used in Models C and E.  This may be one reason why these models
underproduce the local luminsoity density and EBL, since stars with
$\lesssim 2M_\odot$ dominantly produce $\lesssim 2$~eV photons.  The
other affecting factors are the differences in SFR models (see
Fig.~\ref{fig:sfr}) and mass-to-light ratio.  With $f_L >1$ in
equation~(\ref{mass-luminosity}), for the Baldry-Glazebrook IMF, it
may be possible to close gaps between Models C and E with other models
with BPL $L-M$ relation.  However, we do not explore that possibility
in the present paper.  For the purposes of $\gamma$-ray astronomy, any
of the Models A, B or D can reproduce local UV/optical EBL data.  Next
we compare our EBL models to the models by other authors.

\subsection{Comparisons with other authors}
\label{sec:comparisons}

\begin{figure}
\plotone{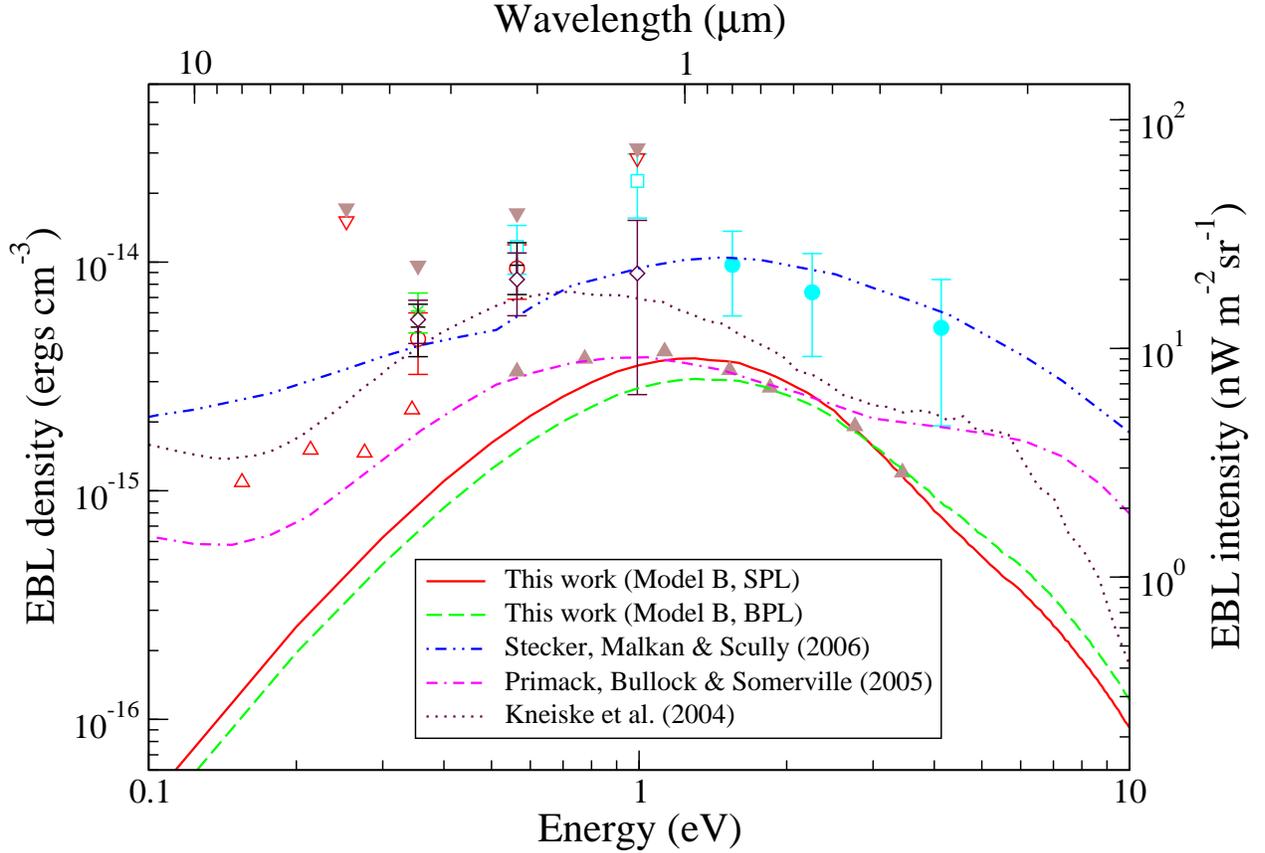}
\caption{A comparison of the EBL models (B, SPL) and (B, BPL) derived
  in this paper with models by other authors.  The updated fast
  evolution model by Stecker, Malkan \& Scully (2006) is based on
  backward-evolution models of local galaxies.  The models by Primack,
  Bullock \& Somerville (2005) is based on Monte Carlo simulatons of
  galaxy evolution with initial conditions.  The model by Kneiske et
  al. (2004) is based on results of population synthesis models.  Our
  models are based on initial mass function and star formation models
  convolved with stellar properties. }
\label{fig:EBL_compare}
\end{figure}

Among several existing models of the EBL, the one by Primack, Bullock
\& Somerville (2005), plotted in Fig.~\ref{fig:EBL_compare} is used to
be considered as the ``Low'' EBL model.  It is consistent with the
lower limits from the galaxy counts above $\sim$0.5~eV.  An updated
version of the fast evolution model by Stecker, Malkan \& Scully
(2006) is also plotted in Fig.~\ref{fig:EBL_compare}.  This is
considered to be the ``High'' EBL model.  The best-fit model by
Kneiske et al. (2004), plotted here, and the phenomenological fits by
Dermer (2007), not plotted here, are generally in between these two
models.

Primack, Bullock \& Somerville (2005) used Monte Carlo simulations of
galaxy evolution and emission by the evolving galaxy population to
calculate their EBL model.  Kneiske, Mannheim \& Hartmann (2002); and
Kneiske et al. (2004) used the outcome of the simulations of stellar
population luminosity at different redshift, and integrated over
redshift to calculate the EBL.  Stecker, Malkan \& Scully (2006)
assumed that the observed luminosity of a galaxy at 60~$\mu m$ can be
used to calculate its luminosity at all wavelengths, and used galaxy
luminosity functions to calculate the EBL.  In our models, we used
blackbody emission from stars, based on the solar models, and convolve
with the star formation and initial mass function models to calculate
the EBL.  These methods, therefore, are not directly comparable to
each other.  However, assumptions about an SFR or/and an IMF model(s)
should affect all methods as also shown by Kneiske, Mannheim \&
Hartmann (2002); Kneiske et al. (2004).  Indeed the variations between
our EBL models A--E results from the differences in SFR and IMF
combinations.

Our models (A, SPL), (B, SPL) and (D, SPL) are consistent with the
{\em HST} lower limits above $\sim$1~eV and agree with the EBL model
by Primack, Bullock \& Somerville (2005) in the $\sim$1--3~eV range.
At energies lower than $\sim$1~eV our models are lower than all other
models.  This is probably because the infrared EBL component, from
dust radiation; luminous infrared galaxies and post main-sequence
stars, becomes important at these lower energies.  At energies above
$\sim$3~eV all other EBL models are higher than our models.  There may
be additional contributions from AGN and white dwarfs at these higher
energies, however due to a lack of data points comparisons between the
models above $\sim$3~eV become less meaningful.

Both the models (B, SPL) and (D, SPL) better represent the luminosity
density data in Fig.~\ref{fig:lumdensity} than the model (A, SPL) and
can be used as our best-fit EBL models. For illustration and further
calculations, we have chosen to use Model B, SPL plotted in
Fig.~\ref{fig:EBL_compare}.  We have also plotted our Model (B, BPL)
for comparisions.

\subsection{Evolution with redshift}
\label{sec:evolution}

\begin{figure}
\plotone{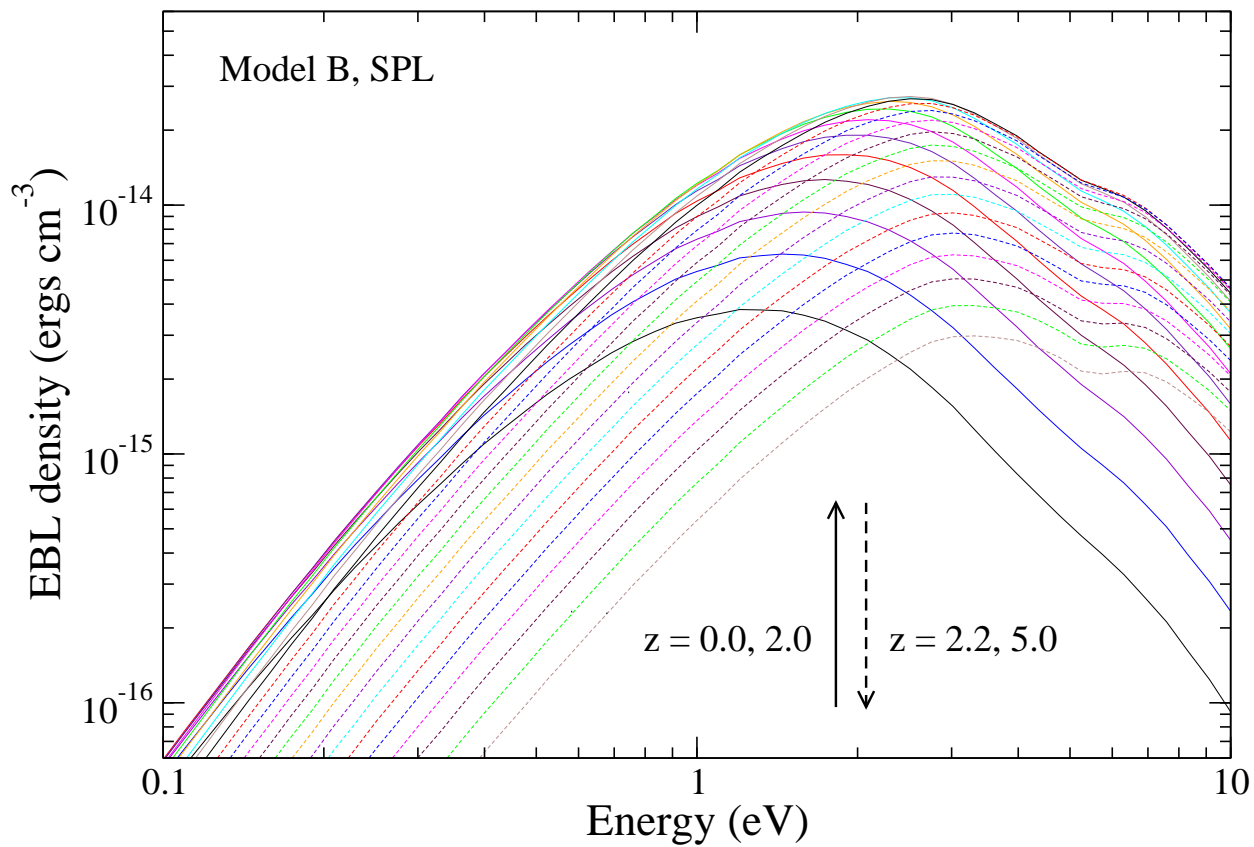}
\caption{ Evolution of the comoving EBL energy density $\eps_1
u_{\eps_1}$ as a function of the comoving photon energy $\eps_1$ with
redshift for our best-fit EBL model (Model B with SPL $L-M$ relation).
Model (D, SPL) gives simililar results.  The curves are plotted for
$z=0$--5 with 0.2 interval.  The density increases first and then
decreases as indicated by the solid and dashed arrows.}
\label{fig:EBL_z} 
\end{figure}

Understanding the evolution of the background light with redshift is a
key to $\gamma$-ray astronomy, which we discuss in the next section.
The mean-free-path for $\gamma\gamma$ absorption with the EBL spans
astronomical distances, over which the EBL itself changes noticeably.
We have plotted the comoving EBL energy density at different redshifts
0--5 in Fig.~\ref{fig:EBL_z} using equation~(\ref{transformations})
for the Model B, SPL.  Initially the EBL density increases with
redshift because of the sharp rise in star formation (see
Fig.~\ref{fig:sfr}) below $z\sim 2$ and a decreasing volume.  At
redshift $\gtrsim 2$, the EBL density decreases because the total
number of stars formed up to that redshift from $z=6$ decreases.  The
contribution to the EBL above $\sim$2--3~eV dominantly comes from high
mass stars.  Since the livetimes of high-mass stars are shorter, their
contribution to the EBL at a particular $z$ is mostly determined by
how many of them are formed at $\sim z$ (see Fig.~\ref{fig:zdeath}).
On the other hand the overall population of low mass stars, which
dominantly produce lower energy photons, increases with decreasing
redshift.  As a result, the ratio of high-energy photon density
increases with redshift (as evidenced by the hump at $\sim$7~eV)
compared to the low energy photon density reflecting a decreasing
overall population of low mass stars for $z\to 6$.

\section{Implications for $\gamma$-ray Astronomy}
\label{sec:implications}

High-energy photons in the $\sim$10--300~GeV energy range from sources
such as GRB at $z\gtrsim 0.5$ are subject to $\gamma\gamma \to e^+e^-$
absorption by the EBL starlight photons as we modeled here.  To
calculate the $\gamma\gamma$ absorption opacity one needs to take into
account an evolving EBL with redshift as plotted in
Fig.~\ref{fig:EBL_z}.  We calculate this opacity both numerically,
using interpolation to the exact results from Model B, SPL
calculation, and analytically, by fitting the Model B, SPL results.

\begin{figure}
\plotone{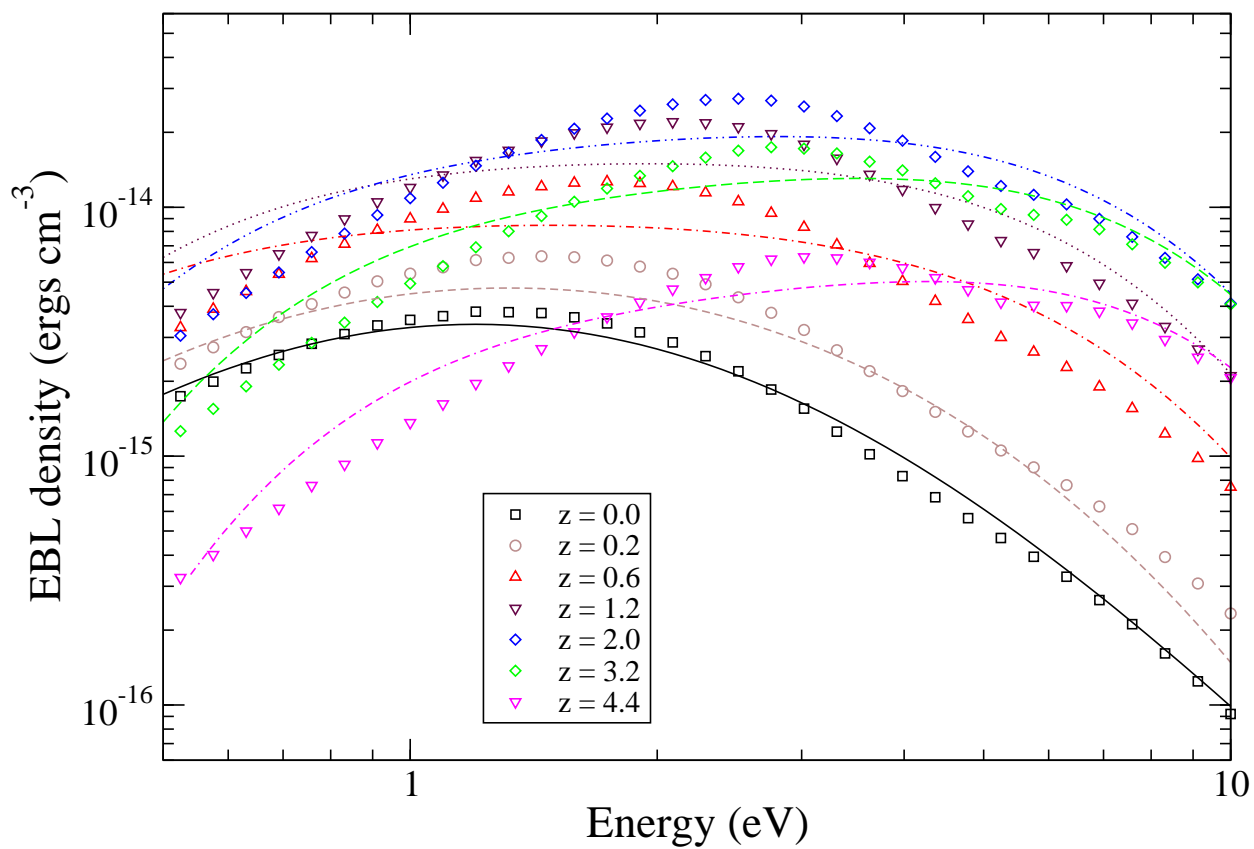}
\caption{Fits (smooth lines) to an evolving EBL energy density $\eps_1
u_{\eps_1}$ (points) as a function of the comoving frame photon energy
$\eps_1$ at different redshifts for Model (B, SPL). The fit function
is given in equation~(\ref{EBL_fit}). } 
\label{fig:EBL_fit_z}
\end{figure}

We provide such a fit below and plot it in Fig.~\ref{fig:EBL_fit_z} at
different redshift overlaid with numerical calculation.  The
polynomial fit parameters are
\ba 
{\rm log} (\eps_1 u_{\eps_1} /{\rm ergs~cm}^{-3}) = 
A_0 + A_1x + A_2x^2 +A_3x^3 + A_4x^4
\nonumber \\
x = {\rm log} (\eps_1~({\rm eV})/(1+z_1)) \nonumber \\
A_0 = -14.4829 + 0.8275 z_1 - 0.2451 z_1^2 + 0.0046 z_1^3 +
 0.0043 z_1^4 - 0.0004 z_1^5 \nonumber \\
A_1 = 0.3157 - 1.105 z_1 + 1.1026 z_1^2 - 0.4764 z_1^3 +
 0.09 z_1^4 - 0.0062 z_1^5 \nonumber \\
A_2 = \cases{ -1.9888 + 1.6527 z_1 + 1.0294 z_1^2 ~;~ z_1<0.8 \cr 
 -0.5549 - 0.0295 z_1 - 0.1133 z_1^2 + 0.0079 z_1^3 ~;~ z_1>0.8 } 
\nonumber \\
A_3 = -0.1507 + 0.9114 z_1 - 1.8907 z_1^2 + 0.8816 z_1^3
 - 0.1837 z_1^4 + 0.0141 z_1^5 \nonumber \\
A_4 = 0.3014 - 3.5371 z_1 + 2.4574 z_1^2 - 0.8474 z_1^3 +
 0.1343 z_1^4 - 0.0079 z_1^5 ~,
\label{EBL_fit}
\ea
where $\eps_1 = \eps(1+z_1)$.  The fit is generally good at the tens
of percentage level of the numerical results for $\eps_1 \gtrsim
0.5$~eV for all redshift $z_1 \le 5$.

With a fit to the diffuse background photons, in
equation~(\ref{EBL_fit}), we can easily calculate the optical depth of
$\gamma \gamma$ absorption for an energetic photon originating at
redshift $z$ with observed energy $E$ as (Gould \& Schr\'eder 1967;
Brown, Mikaelian \& Gould 1973)
\ba
\tau_{\gamma\gamma} (E, z) &=& 
c\int_0^z dz_1 \left| \frac{dt}{dz_1} \right| 
\nonumber \\ && ~ \times
\int_{0}^{\infty}  d\eps_1 \int_{-1}^{1} d\cos\theta 
\frac{1}{2} \frac{u_{\eps_1}}{\eps_1} (1-\cos\theta) 
\sigma_{\gamma\gamma} (s) ~~
\nonumber \\ &=& 
c\pi r_e^2 \frac{m_e^4c^8}{E^2}
\int_0^z \frac{dz_1}{(1+z_1)^2} \left| \frac{dt}{dz_1} \right| 
\nonumber \\ && ~\times
\int_{m_e^2 c^4/E(1+z_1)}^{\infty} {d\eps_1} 
\frac{u_{\eps_1}}{\eps_1} {\bar \varphi} [s_0 (\eps_1)].
\label{gg_opacity}
\ea
for an isotropic background photon field.  Here $\sigma_{\gamma\gamma}
(s)$ is the total $\gamma\gamma \to e^+ e^-$ cross-section (see, e.g.,
Jauch \& Rohrlich 1955) and $s = E(1+z_1) \eps_1
(1-\cos\theta)/2m_e^2c^4$ is the center-of-mass energy squared.  The
function ${\bar \varphi} [s_0]$, with $s_0 = E(1+z_1) \eps_1
/m_e^2c^4$, is given in Gould \& Schr\'eder (1967).  The threshold
energy for $e^+e^-$ pair production from the condition $s_0=1$ is
$\eps_{1,\rm th} \approx m_e^2c^4/E(1+z_1)$.  The lower limit of the
energy integration $\eps_{1,\rm th} \approx 1$~eV and $\sim
250/(1+z_1)$~GeV $\gamma$-rays interact dominantly with these photons.
Higher energy $\gamma$-rays mainly interact with softer photons, at an
energy range where the re-processed dust emission dominates the EBL
and our models do not fit the data.  With an upper limit $\eps_{1,\rm
max} \approx 10$~eV, it is safe to use equation~(\ref{gg_opacity}) in
the $\sim$10--300~GeV $\gamma$-ray energy range for redshift $\gtrsim
0.5$.  Note that this the relevant energy range for the Large Area
Telescope on board the {\em Fermi Gamma Ray Space Telescope.}  Air
Cherenkov Telescopes also become sensitive to $\gamma$-rays
$\gtrsim$50~GeV.  The survival probability for a $\gamma$-ray created
at redshift $z$ to reach Earth is $\exp[-\tau_{\gamma\gamma} (E,z)]$.
The observed flux of $\gamma$-rays is attenuated by this factor from
the source flux.

\begin{figure}
\plotone{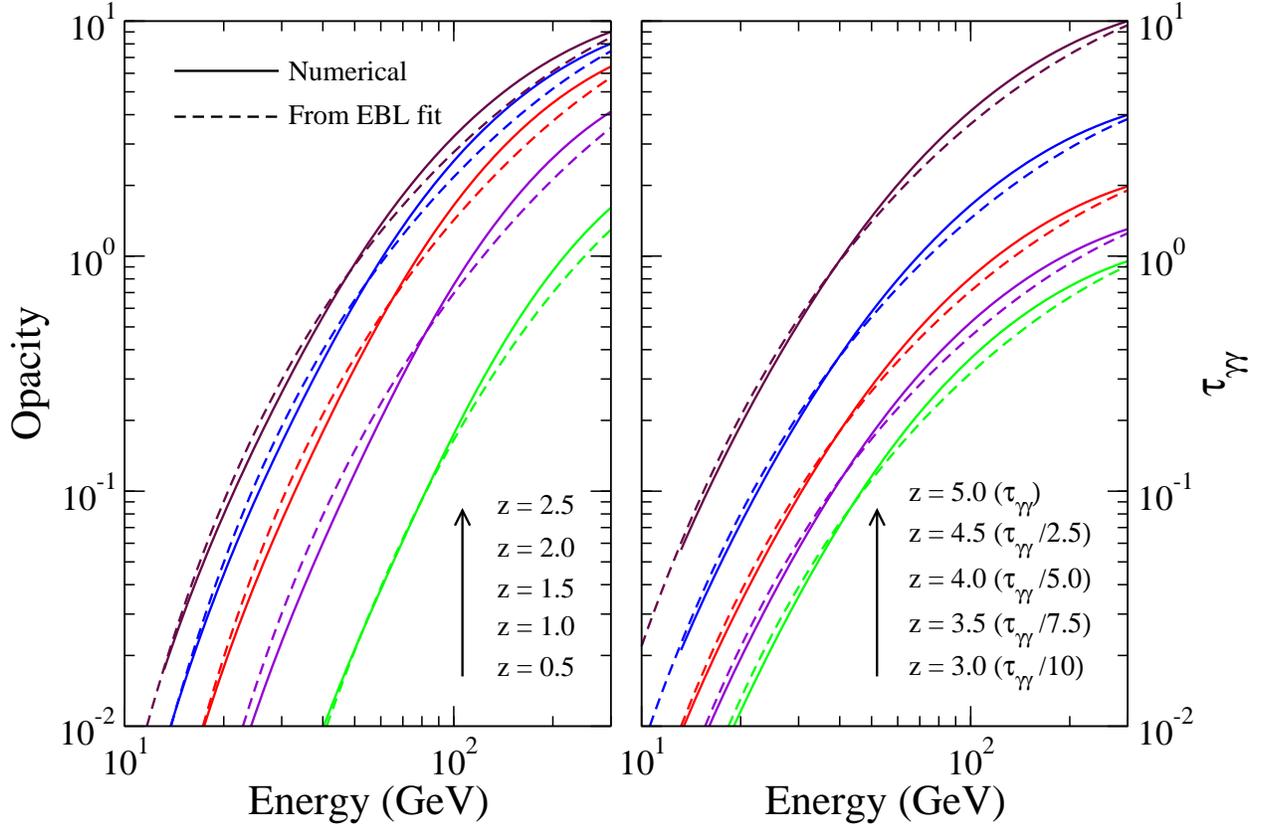}
\caption{Gamma-ray opacity of the universe at different redshifts ({\em
Left panel:} 0.5--2.5, {\em Right panel: 3.0--5.0}) and energies as
plotted here.  The solid and dashed lines correspond to the exact
calculation and calculation using the fit to the EBL in
equation~\ref{EBL_fit}, respectively.  We rescaled
$\tau_{\gamma\gamma}$ in the {\em Right panel} to separate diferent
curves.  The numerical values for the exact calculation are also
listed in Table~\ref{tab:gg_opacity}.}  
\label{fig:tau_gg}
\end{figure}

\begin{figure}
\plotone{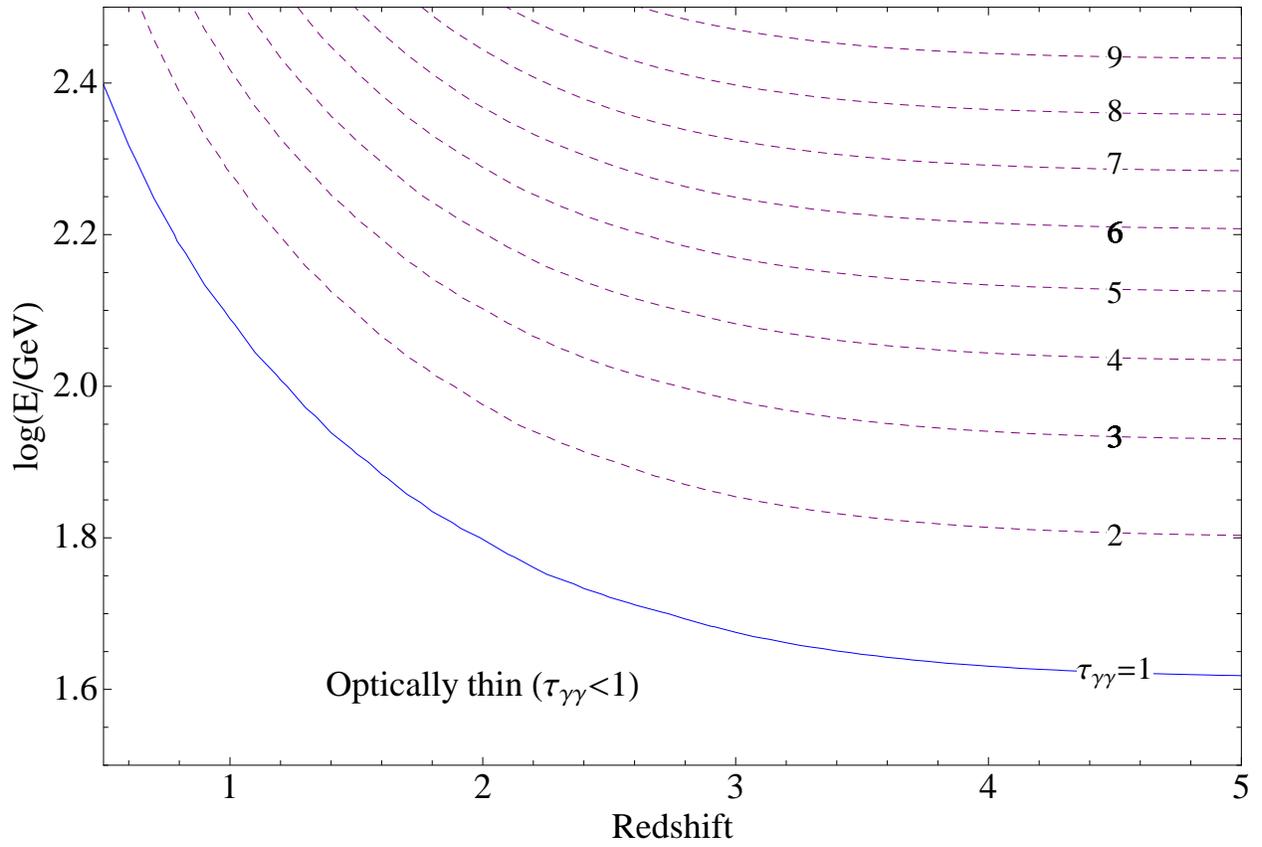}
\caption{Gamma ray opacity contrours in the $E-z$ plane using the EBL
fit in equation~(\ref{EBL_fit}).  The $\tau_{\gamma\gamma} =1$ contour
plotted here is known as the Fazio-Stecker relation (Fazio \&
Stecker~1970) and represents a $\gamma$-ray horizon of the universe.}
\label{fig:opt_contour} 
\end{figure}

We have plotted the $\gamma\gamma$ opacity ($\tau_{\gamma\gamma}$) in
Fig.~\ref{fig:tau_gg} for redshift 0.5--5.0 and observed $\gamma$-ray
energy 10--300~GeV using equation~(\ref{gg_opacity}).  The solid and
dashed lines correspond to the exact calculation and calculation using
the fit in equation~(\ref{EBL_fit}) respectively.  As can be seen, the
two methods of calculation give results within tens of percent.  For
$z\ge 3$ (right panel) we have divided $\tau_{\gamma\gamma}$ with
constant factors given on the plot to avoid cluttering of different
curves. We also provide our exact calculation of $\tau_{\gamma\gamma}$
in Table~\ref{tab:gg_opacity} for different redshift and $\gamma$-ray
energies.  The relation $\tau_{\gamma\gamma} =1$
(Fig.~\ref{fig:opt_contour}), sometimes referred to as the
Fazio-Stecker relation (Fazio \& Stecker 1970), corresponds to a
$\gamma$-ray horizon of the universe.

\begin{figure} 
\plotone{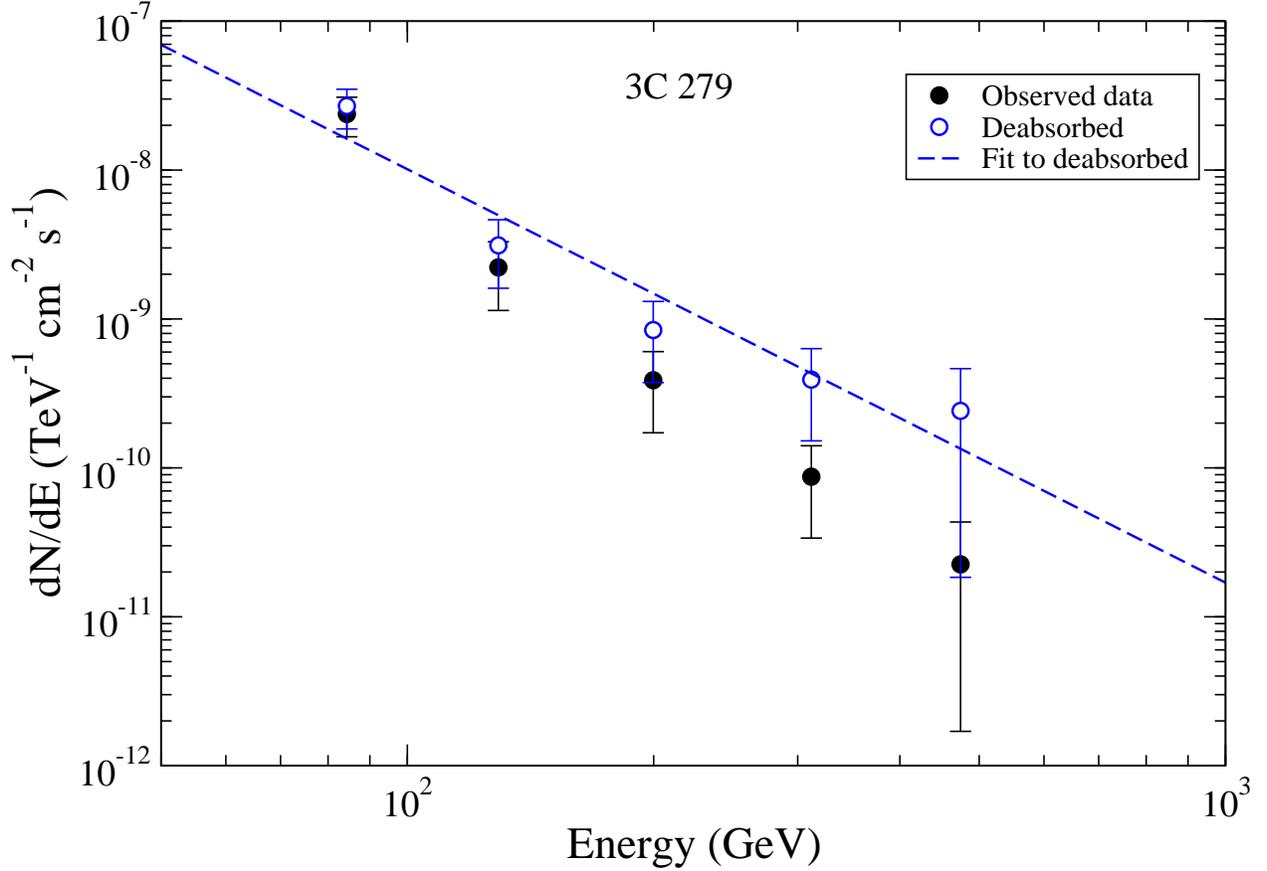} 
\caption{ High-energy $\gamma$-ray flux from the furthest known
($z=0.536$) blazar 3C 279 as measured by MAGIC (filled circles with
errorbars.  We use our EBL Model B, SPL to estimate (deabsorb) the
source flux (open circles with errorbars) by multiplying the observed
data points with $\exp(\tau_{\gamma\gamma})$.  The dashed line is a
power-law fit $dN/dE \propto E^{-\Gamma}$ to the deabsrbed data
points resulting in a spectral index $\Gamma = 2.8$.}
\label{fig:blazar_deabsorbed} 
\end{figure}

With the calculated opacities, we can calculate the true $\gamma$-ray
flux of a distant source before absorption in the EBL by multiplying
the observed flux $f_E$ with the factor
$\exp[\tau_{\gamma\gamma}(E)]$. In case of the furthest known blazar
3C279 at $z=0.536$, the flux data points measured by MAGIC (Albert et
al. 2008) are plotted in Fig.~\ref{fig:blazar_deabsorbed}. We have
also plotted the deabsorbed data points (open circles) using our EBL
Model B (SPL) and a power-law fit $dN/dE = N_0 (E/200~{\rm
GeV})^{-\Gamma}$ with $N_0 = 1.5\times
10^{-9}$~TeV$^{-1}$~cm$^{-2}$~s$^{-1}$ and $\Gamma = 2.8$.  The
deabsorbed spectrum of 3C 279 thus becomes harder than the observed
value of $\Gamma = 4.11$ (Albert et al. 2008).  Note, however, that
our $\gamma\gamma$ opacity calculation is less reliable for the
highest energy data point at 474~GeV. The $\Gamma$ for the deabsorbed
spectrum, thus, can be $\lesssim 2.78$ if $\tau_{\gamma\gamma}$ is much
larger than our estimate.

Simple Fermi mechanisms of particle acceleration in both relativitic
and non-relativistic shocks, result in particle spectra $dN/dE \propto
E^{-p}$ with $p\approx 2$.  In the case of a one-zone synchro-Compton
mechanism to produce TeV $\gamma$-rays at the shocks, the source
spectrum of $\gamma$-rays typically should be larger than $\Gamma =
(p+1)/2 \approx 1.5$.  Spectra harder than the simple-minded limiting
value may arise in a number of scenarios.  For a special environment
at or surrounding the shocks, effects such as a cutoff of
shock-accelerated elecron spectrum below a very high energy $\sim$GeV
(Katarzy\'nski et al. 2006; Stern \& Poutanen 2008) or internal
$\gamma\gamma$ absorption in a dense region of quasi mono-energetic
soft photons (Aharonian, Khangulyan \& Costamante 2008) may result in
a very hard spectrum $\Gamma < 1.5$ over a short energy range.
Acceleration at shear flows (Stawarz \& Ostrowski 2002; Rieger \&
Duffy 2006) or Monte Carlo shock-acceleration models may also result
in $\Gamma < 1.5$ (Virtanen \& Vainio 2005; Stecker et al. 2007).

Models with multiple spectral components, arising from multiple zones,
are attractive to explain observed hard blazar spectra.  Recently
B\"ottcher, Dermer \& Finke (2008) have invoked such a model of
Compton upscattering of CMB photons in large scale jet to explain hard
spectrum of 1ES 1101-232.  Also, cosmic-ray acceleration and
interactions at the shocks and subsequent cascade in the EBL-CMB may
also produce very hard $\gamma$-ray spectra (Coppi \& Aharonian 1997).
However, such models cannot explain highly variable TeV spectra, e.g.,
from 3C 279 (Albert et al. 2008).

\section{Conclusions}
\label{sec:conclusions}

We have derived a class of well-defined models for the spectral energy
density of the EBL.  The models with modified Salpeter A initial mass
function, a single power-law mass-luminosity relation and Cole et
al. (2001) or Hopkins \& Beacom (2006) star formation history
reasonably fit the EBL UV-optical data and the luminosity density in
our local universe.

Our models are based on the underlying assumption that the bulk of the
EBL radiation between $\sim 0.1$ -- 10 eV is due to stellar radiations
absorbed by dust, which can be determined from recent analyses of
galaxies by Driver et al.\ (2008).  This approach differs from models
by Stecker, Malkan \& Scully (2006) based on luminosity evolution of
galactic spectral energy distributions, which is limited to the
accuracy of the available observational data used in the survey.
Without need of a population synthesis code (e.g., models by Primack,
Bullock \& Somerville 2005) or fits to the results of such a code
(e.g., models by Kneiske, Mannheim \& Hartmann 2002), our model is
based on well-studied results from stellar astronomy.

The sources of uncertainties in our models are the (i) main-sequence
age of the star and stellar luminosity, discussed earlier, (ii) star
formation history and initial mass function, and (iii) dust
absorption.  We have already discussed point (ii) in some details
using five SFR+IMF models.  Note that in all those models, the IMF was
assumed to be independent of redshift.  In principle the normalization
of the IMF or even its shape may depend on $z$.  Nevertheless, a
universal IMF fits SFR data reasonably well.  The evolution of the
dust absorption model with redshift, which we have not taken into
account, is of potentially greater concern.  At high redshift it is
more reasonable to assume that the dust absorption (escape) fraction
would be higher (lower), so that the stellar contribution to the EBL
at high $z$ would be less than if using a constant absorption
fraction.  A more detailed examination of these issues are under
further study (Finke et al., in perperation).

We have provided an analytic fit to our best-fit EBL model and its
evolution with redshift.  This result can be used to calculate, as we
have done in this work, opacity of the universe to $\sim$10--300~GeV
$\gamma$-rays relevant for the high energy data from the {\em Fermi
Gamma Ray Space Telescope} and Air Cherenkov Telescopes, and
estimating unknown quantities such as the spectrum and energy at
production of the distant GRBs and blazars such as 3C 279.

\acknowledgements We thank Eli Dwek, Claude-Andr{\' e}
Faucher-Gigu{\`e}re, Dieter Hartmann, Tanja Kneiske, Kalevi Mattila
and Floyd Stecker for helpful comments.  The work of S.R. and
J.D.F. was supported by the National Research Council Research
associateship program at the Naval Research Laboratory.  The work of
C.D.D. was supported by the Office of Naval Research.

\clearpage

\begin{deluxetable}{lcccccccccc}
\tablewidth{0pt}
\tablecolumns{11}
\tablecaption{\label{tab:gg_opacity}
$\gamma$ ray absorption opacity ($\tau_{\gamma\gamma}$) with EBL Model B, SPL
at different redshift }
\tablehead{
\colhead{$E_\gamma$} &
\colhead{$z$} &
\colhead{$z$} &
\colhead{$z$} &
\colhead{$z$} &
\colhead{$z$} &
\colhead{$z$} &
\colhead{$z$} &
\colhead{$z$} &
\colhead{$z$} &
\colhead{$z$} \\
\colhead{GeV} &
\colhead{$0.5$} &
\colhead{$1.0$} &
\colhead{$1.5$} &
\colhead{$2.0$} &
\colhead{$2.5$} &
\colhead{$3.0$} &
\colhead{$3.5$} &
\colhead{$4.0$} &
\colhead{$4.5$} &
\colhead{$5.0$} }
\startdata
 13.18 & 0.0000 & 0.0004 & 0.0025 & 0.0081 & 0.0174 & 0.0284 & 0.0387 & 0.0468 & 0.0526 & 0.0563 \\
 14.45 & 0.0000 & 0.0007 & 0.0041 & 0.0123 & 0.0253 & 0.0400 & 0.0533 & 0.0635 & 0.0708 & 0.0752 \\
 15.85 & 0.0001 & 0.0013 & 0.0065 & 0.0183 & 0.0359 & 0.0551 & 0.0721 & 0.0849 & 0.0938 & 0.0993 \\
 17.38 & 0.0002 & 0.0021 & 0.0099 & 0.0264 & 0.0498 & 0.0747 & 0.0961 & 0.1121 & 0.1229 & 0.1294 \\
 19.05 & 0.0003 & 0.0034 & 0.0146 & 0.0372 & 0.0680 & 0.0998 & 0.1266 & 0.1462 & 0.1594 & 0.1670 \\
 20.89 & 0.0005 & 0.0053 & 0.0212 & 0.0515 & 0.0914 & 0.1316 & 0.1649 & 0.1889 & 0.2046 & 0.2136 \\
 22.91 & 0.0008 & 0.0079 & 0.0300 & 0.0700 & 0.1213 & 0.1718 & 0.2127 & 0.2418 & 0.2604 & 0.2708 \\
 25.12 & 0.0014 & 0.0116 & 0.0416 & 0.0940 & 0.1592 & 0.2221 & 0.2722 & 0.3068 & 0.3286 & 0.3405 \\
 27.54 & 0.0021 & 0.0167 & 0.0568 & 0.1246 & 0.2070 & 0.2849 & 0.3452 & 0.3861 & 0.4113 & 0.4249 \\
 30.20 & 0.0032 & 0.0234 & 0.0765 & 0.1636 & 0.2671 & 0.3622 & 0.4342 & 0.4818 & 0.5108 & 0.5258 \\
 33.11 & 0.0047 & 0.0323 & 0.1019 & 0.2131 & 0.3417 & 0.4566 & 0.5415 & 0.5965 & 0.6292 & 0.6457 \\
 36.31 & 0.0068 & 0.0440 & 0.1345 & 0.2755 & 0.4335 & 0.5709 & 0.6698 & 0.7324 & 0.7687 & 0.7869 \\
 39.81 & 0.0096 & 0.0592 & 0.1763 & 0.3533 & 0.5454 & 0.7078 & 0.8215 & 0.8916 & 0.9317 & 0.9515 \\
 43.65 & 0.0134 & 0.0789 & 0.2295 & 0.4493 & 0.6804 & 0.8696 & 0.9984 & 1.0763 & 1.1203 & 1.1410 \\
 47.86 & 0.0183 & 0.1045 & 0.2963 & 0.5667 & 0.8410 & 1.0582 & 1.2027 & 1.2883 & 1.3357 & 1.3580 \\
 52.48 & 0.0248 & 0.1375 & 0.3797 & 0.7085 & 1.0296 & 1.2761 & 1.4361 & 1.5293 & 1.5798 & 1.6037 \\
 57.54 & 0.0334 & 0.1798 & 0.4828 & 0.8773 & 1.2479 & 1.5247 & 1.7004 & 1.8002 & 1.8545 & 1.8788 \\
 63.10 & 0.0446 & 0.2335 & 0.6084 & 1.0751 & 1.4978 & 1.8049 & 1.9955 & 2.1019 & 2.1592 & 2.1844 \\
 69.18 & 0.0593 & 0.3010 & 0.7590 & 1.3034 & 1.7795 & 2.1166 & 2.3210 & 2.4340 & 2.4933 & 2.5199 \\
 75.86 & 0.0785 & 0.3848 & 0.9365 & 1.5636 & 2.0939 & 2.4597 & 2.6772 & 2.7958 & 2.8578 & 2.8840 \\
 83.18 & 0.1033 & 0.4871 & 1.1428 & 1.8562 & 2.4409 & 2.8337 & 3.0635 & 3.1867 & 3.2510 & 3.2783 \\
 91.20 & 0.1349 & 0.6102 & 1.3792 & 2.1814 & 2.8190 & 3.2373 & 3.4784 & 3.6056 & 3.6714 & 3.6997 \\
100.00 & 0.1750 & 0.7558 & 1.6461 & 2.5388 & 3.2267 & 3.6688 & 3.9199 & 4.0500 & 4.1189 & 4.1456 \\
109.65 & 0.2250 & 0.9258 & 1.9440 & 2.9260 & 3.6617 & 4.1259 & 4.3848 & 4.5181 & 4.5881 & 4.6164 \\
120.23 & 0.2862 & 1.1201 & 2.2714 & 3.3401 & 4.1212 & 4.6048 & 4.8703 & 5.0067 & 5.0768 & 5.1051 \\
131.83 & 0.3596 & 1.3394 & 2.6262 & 3.7782 & 4.6013 & 5.1010 & 5.3727 & 5.5094 & 5.5818 & 5.6089 \\
144.54 & 0.4466 & 1.5839 & 3.0055 & 4.2369 & 5.0978 & 5.6109 & 5.8874 & 6.0243 & 6.0973 & 6.1244 \\
158.49 & 0.5473 & 1.8526 & 3.4069 & 4.7131 & 5.6059 & 6.1303 & 6.4087 & 6.5471 & 6.6193 & 6.6479 \\
173.78 & 0.6626 & 2.1416 & 3.8259 & 5.1994 & 6.1200 & 6.6530 & 6.9323 & 7.0694 & 7.1430 & 7.1695 \\
190.55 & 0.7923 & 2.4489 & 4.2581 & 5.6910 & 6.6349 & 7.1731 & 7.4531 & 7.5872 & 7.6624 & 7.6876 \\
208.93 & 0.9363 & 2.7721 & 4.6986 & 6.1836 & 7.1457 & 7.6864 & 7.9641 & 8.0971 & 8.1708 & 8.1969 \\
229.09 & 1.0925 & 3.1074 & 5.1405 & 6.6710 & 7.6465 & 8.1867 & 8.4607 & 8.5922 & 8.6658 & 8.6900 \\
251.19 & 1.2597 & 3.4505 & 5.5803 & 7.1482 & 8.1314 & 8.6679 & 8.9390 & 9.0668 & 9.1404 & 9.1650 \\
275.42 & 1.4362 & 3.7960 & 6.0133 & 7.6094 & 8.5944 & 9.1251 & 9.3916 & 9.5161 & 9.5892 & 9.6121 \\
301.99 & 1.6204 & 4.1397 & 6.4326 & 8.0473 & 9.0297 & 9.5530 & 9.8138 & 9.9358 & 10.007 & 10.031
\enddata
\end{deluxetable}

\clearpage

\begin{figure}
\plotone{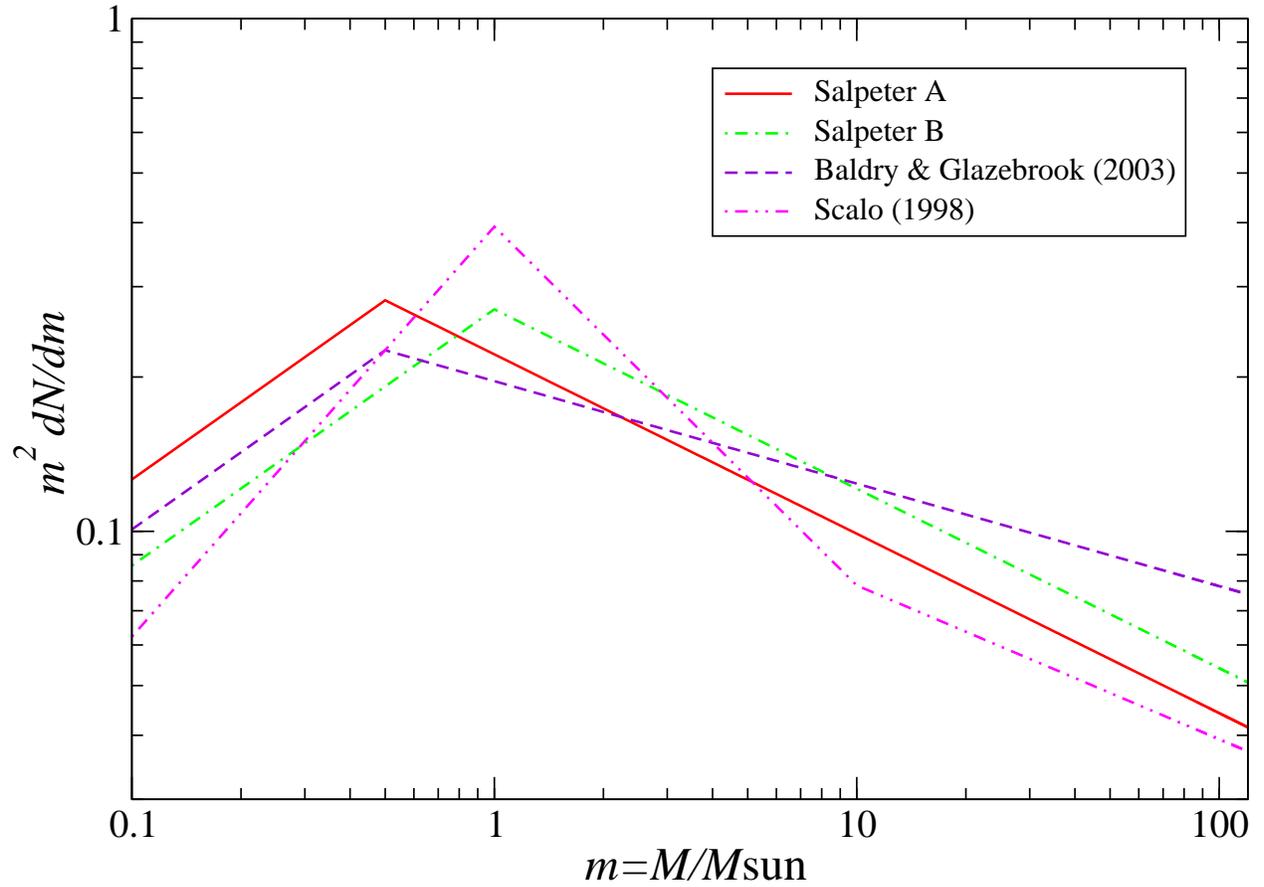}
\caption{Initial mass function (IMF) models, assumed to be
  universal. The integral of $dN/d{\rm ln}M$ is set to unity over
  $M$=(0.1--120)$M_\odot$ range.}
\label{fig:imf}
\end{figure}

\end{document}